 \documentclass[final,onecolumn,3p,times,sort&compress,12pt]{elsarticle}
 \makeatletter
\def\ps@pprintTitle{%
 \let\@oddhead\@empty
 \let\@evenhead\@empty
 \def\@oddfoot{\centerline{\thepage}}%
 \let\@evenfoot\@oddfoot}
\makeatother

\usepackage{amssymb}
\usepackage[breaklinks=true]{hyperref}
\usepackage{breakurl,epstopdf,xcolor}

\journal{Journal of Magnetism and Magnetic Materials}

	\usepackage{fancyheadings}
\renewcommand{\thispagestyle}[1]{} 

\begin{document}



\title{$\,$\\Ground-state magnetic properties of spin ladder-shaped quantum nanomagnet: Exact diagonalization study}


\author[a1]{K. Sza\l{}owski\corref{cor1}}
\ead{kszalowski@uni.lodz.pl}
\author[a1]{P. Kowalewska}
\address[a1]{Department of Solid State Physics, Faculty of Physics and Applied Informatics,\\
University of \L\'{o}d\'{z}, ulica Pomorska 149/153, 90-236 \L\'{o}d\'{z}, Poland}

\cortext[cor1]{Corresponding author}

\date{\today}

\begin{abstract}
The paper presents a computational study of the ground-state properties of a quantum nanomagnet possessing the shape of a finite two-legged ladder composed of 12 spins $S=1/2$. The system is described with isotropic quantum Heisenberg model with nearest-neighbour interleg and intraleg interactions supplemented with diagonal interleg coupling between next nearest neighbours. All the couplings can take arbitrary values. The description of the ground state is based on the exact numerical diagonalization of the Hamiltonian. The ground-state phase diagram is constructed and analysed as a function of the interactions and the external magnetic field. The ground-state energy and spin-spin correlations are extensively discussed. The cases of ferro- and antiferromagnetic couplings are compared and contrasted. \\
  \begin{center}
   \includegraphics[width=0.5\columnwidth]{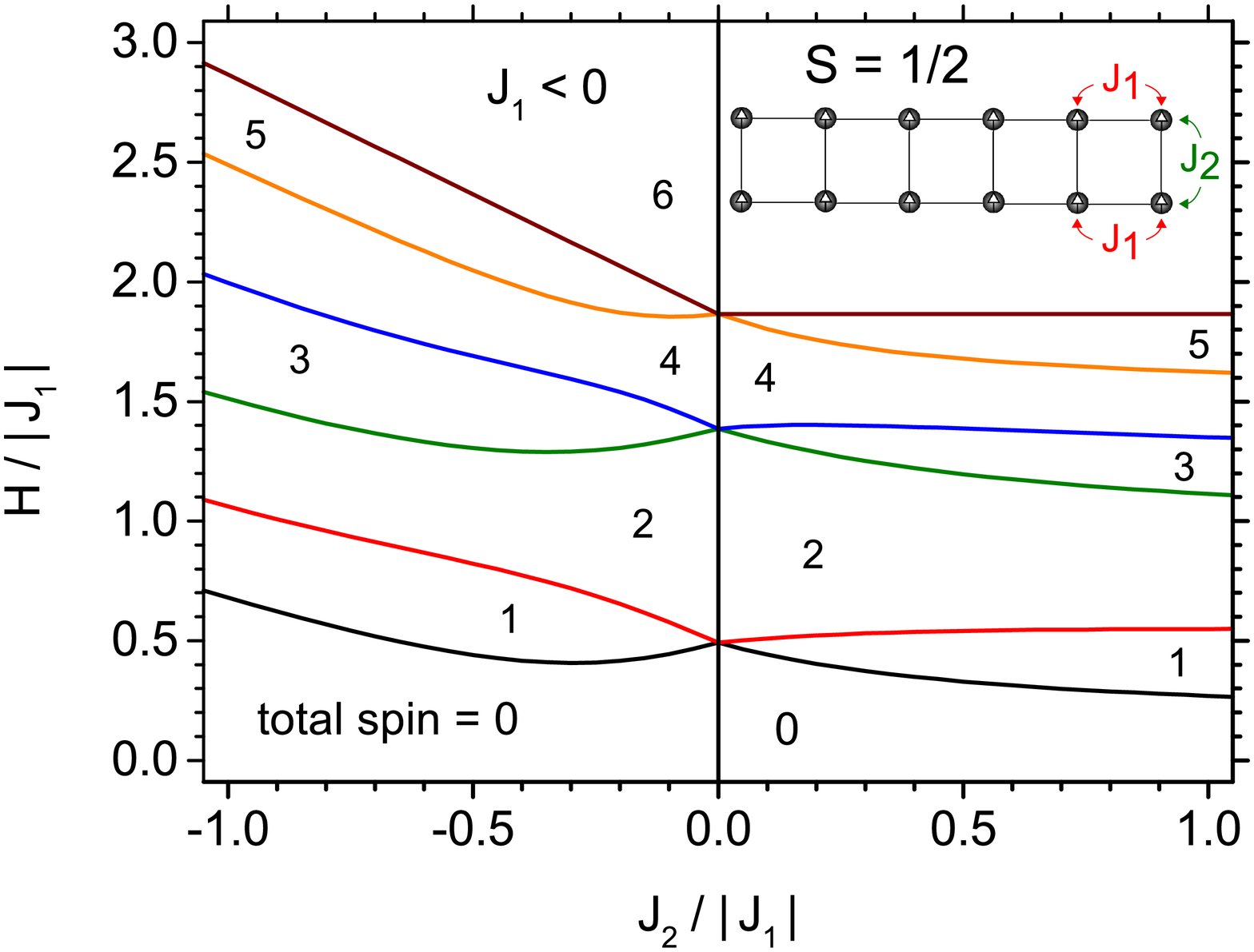}
  \end{center}
 \end{abstract}

\maketitle

\section{Introduction}

Low-dimensional systems attract increasing attention of solid state physicists. The most intensive studies focus on various nanostructures and within this class a considerable attention is paid to magnetic nanosystems \cite{owens_physics_2015,martin_ordered_2003,himpsel_magnetic_1998}.
The principal motivation for studies of the smallest nanostructured magnetic systems is the possibility of arranging them within bottom-up approach from single atoms of surfaces \cite{hirjibehedin_spin_2006,loth_bistability_2012,yan_nonlocally_2017}. This approach allows the design and engineering of artificial nanomagnets with high precision. Moreover, their properties can also be carefully characterized at the nanoscale \cite{ternes_probing_2017,yan_three-dimensional_2015,guidi_direct_2015,holzberger_parity_2013,antkowiak_detection_2013}. What is crucial, the geometry \cite{konstantinidis_design_2013} and underlying magnetic interactions in such systems \cite{florek_sequences_2016,oberg_control_2014,yan_control_2015,delgado_emergence_2015} can be tuned to achieve the desired characteristics. It has been demonstrated that the nanomagnets arranged of single atoms can serve as memory devices, what proves the high potential for applications \cite{yan_nonlocally_2017,loth_bistability_2012}. Moreover, the nanomagnetic systems are also hoped to be useful for quantum computations \cite{leuenberger_quantum_2001}, to mention, for example, spin clusters representing the qubits \cite{meier_quantum_2003,meier_quantum_2003-1}. This route is particularly promising when based on molecular nanomagnets \cite{sieklucka_molecular_2017,friedman_single-molecule_2010,bogani_molecular_2008,wernsdorfer_molecular_2007,schnack_molecular_2004}. The mentioned facts serve as a strong motivation for theoretical studies of a variety of nanomagnetic systems.

One of the interesting classes of such systems is nanomagnets possessing the shape of spin ladder with finite length. This structure was the subject of experimental interest in Ref.~\cite{yan_nonlocally_2017} and built a prototypical memory device. It should be mentioned that major attention in the literature is paid to infinite spin ladders with various number of legs, constituting one-dimensional systems \cite{affleck_quantum_1989,barnes_excitation_1993,cabra_magnetization_1997,honecker_magnetization_2000,mikeska_one-dimensional_2004}. In that context the notion of Haldane gap and the dependence of excitations on spin magnitude and the number of legs in the ladder should be mentioned \cite{affleck_quantum_1989}. However, highly interesting properties can be shown also by the finite systems themselves. Although the magnetic ordering is excluded in such structures, yet they can exhibit interesting magnetic phases and cross-overs between them. Among the studies of such zero-dimensional structures, the works based on exact methods should be especially mentioned \cite{zukovic_magnetization_2017,kariova_enhanced_2017,karlova_magnetization_2017,karlova_isothermal_2017,konstantinidis_influence_2016,zukovic_thermodynamic_2015,strecka_giant_2015,karlova_schottky-type_2016,konstantinidis_antiferromagnetic_2015,zukovic_entropy_2014,
furrer_magnetic_2013,konstantinidis_2009,konstantinidis_antiferromagnetic_2005,schnack_exact_2005,honecker_lanczos_2001,cabra_magnetization_1997}. It is worth emphasizing that rigorous and exact numerical solutions are, so far, available only for a very limited class of models (especially when the quantum version is considered) \cite{braz_quantum_2016,miyahara_exact_2011,strecka_generalized_2010}. 

In order to explore the magnetic properties of the finite structures, it is first vital to examine their ground states, taking into account various possible interactions between the spins as well as the external magnetic field. This is the aim of the present paper, in which we investigate a nanomagnet being a two-legged finite spin ladder with 12 spins $S=1/2$. For this purpose we select an approach based on exact diagonalization, which provides an approximation-free picture of the physics of the studied system. The further parts of the paper contain a detailed description of the system in question, the theoretical approach and the review of the obtained results.

\section{Theoretical model}

The system of interest in the present study is a nanomagnet having the shape of a finite ladder with two (equivalent) legs. The schematic view of the system is presented in Fig.~\ref{fig:1}. It consists of $N=12$ quantum spins $S=1/2$, coupled with isotropic, Heisenberg-like interactions. Therefore, it is described with the following quantum Hamiltonian:

\begin{figure}[ht!]
  \begin{center}
   \includegraphics[width=0.99\columnwidth]{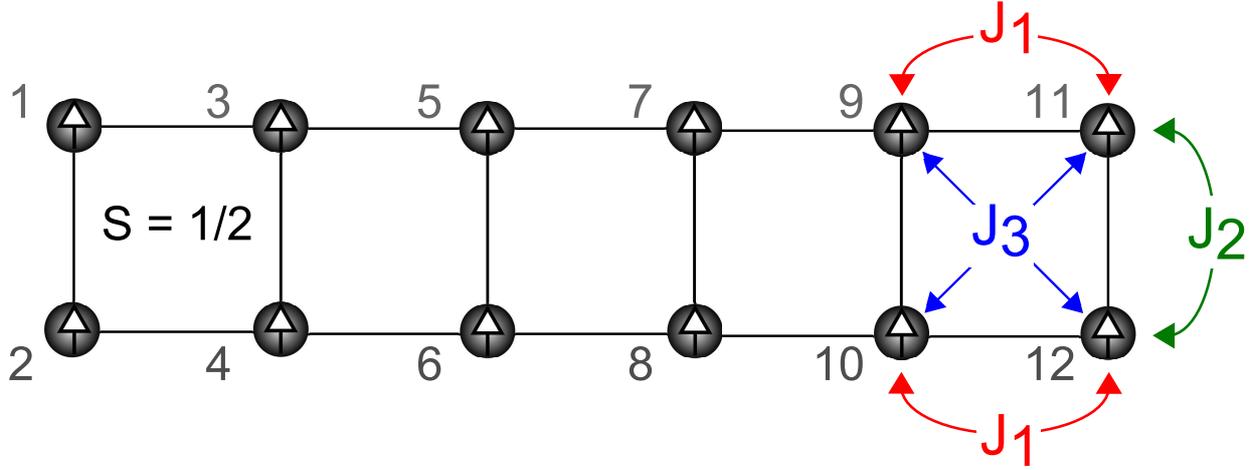}
  \end{center}
   \caption{\label{fig:1} A schematic view of the quantum nanomagnet composed of 12 spins $S=1/2$, having a shape of finite two-legged ladder. The exchange integrals between the spins are depicted schematically. }
\end{figure}

\begin{eqnarray}
\label{eq:1}
\mathcal{H}&=&-J_1\sum_{\left\langle i,j\right\rangle}^{}{\mathbf{S_{i}}\cdot \mathbf{S_{j}}}-J_2\sum_{\left\langle i,j\right\rangle}^{}{\mathbf{S_{i}}\cdot \mathbf{S_{j}}}-J_3\sum_{\left\langle\left\langle i,j\right\rangle\right\rangle}^{}{\mathbf{S_{i}}\cdot \mathbf{S_{j}}}\nonumber\\&&-H\sum_{i}^{}{S^{z}_{i}}. 
\end{eqnarray}
The operator $\mathbf{S_{i}}=\left(S^x_i,S^y_i,S^z_i\right)$ denotes a quantum spin $S=1/2$, located at site labelled with $i$ ($i=1,\dots,12$), with $S_i^{\alpha}=\sigma^{\alpha}/2$, where $\sigma^{\alpha}$ is the appropriate Pauli matrix and $\alpha=x,y,z$ is the direction in spin space. Moreover, the product $\mathbf{S_{i}}\cdot \mathbf{S_{j}}=S^x_iS^x_j+S^y_iS^y_j+S^z_iS^z_j$. The exchange integral between nearest-neighbour spins in the same leg of the ladder amounts to $J_1$, while the interactions between the ladder legs are denoted by $J_2$ for nearest neighbours (rung interactions) and $J_3$ for next-nearest neighbours (crossing interactions); see the scheme Fig.~\ref{fig:1}. All the exchange integrals $J_1,J_2,J_3$ are allowed to take arbitrary values, both positive (ferromagnetic) and negative (antiferromagnetic). The external magnetic field acting in $z$ direction in spin space is introduced by $H$.

In the present study, the interest is focused on the ground-state properties of the system, at zero temperature. In order to perform the description, the full Hamiltonian (Eq.~\ref{eq:1}) is constructed in a form of a matrix (of the size $2^N\times 2^N=4096\times 4096$) and diagonalized numerically \cite{noauthor_mathematica_2010}. This procedure yields the eigenvalues $E_k$ and eigenvectors $\left|\psi_k\right>$ (which can be degenerate). Among the eigenenergies, the ground-state energy $E_0$ is selected, with eigenvectors $\left|\psi_0^p\right>$, where $p=1,\dots,d$ and $d$ is the degeneracy. At the zero temperature, each of the degenerate ground states is equally probable. Therefore, the ground-state average of the arbitrary quantum operator $A$ can be evaluated on the basis of the following formula: $\left<A\right>=\frac{1}{d}\sum_{p=1}^{d}{\left<\psi_0^p\right|A\left|\psi_0^p\right>}$. The observable of special interest is here the total $z$ component of spin of the nanomagnet, with the average value of $S_T=\left\langle\sum_{i=1}^{N}{S^z_i}\right\rangle$. In that context, another quantum number can be defined, namely the total spin quantum number, defined by $\tilde{S}_T\left(\tilde{S}_T+1\right)=\left\langle\sum_{i=1}^{N}{\mathbf{S}^2_i}\right\rangle$. Further important quantities are spin-spin correlation functions $c_{ij}^{\alpha\beta}=\left<S^{\alpha}_iS^{\beta}_{j}\right>$, where $\alpha,\beta=x,y,z$.

The presented theoretical formalism serves as a basis for numerical calculations of the crucial ground-state properties of the studied nanomagnet, which will be discussed in the following section of the paper.

\section{Numerical results and discussion}

\begin{figure*}[h!]
  \begin{center}
   \includegraphics[width=0.9\columnwidth]{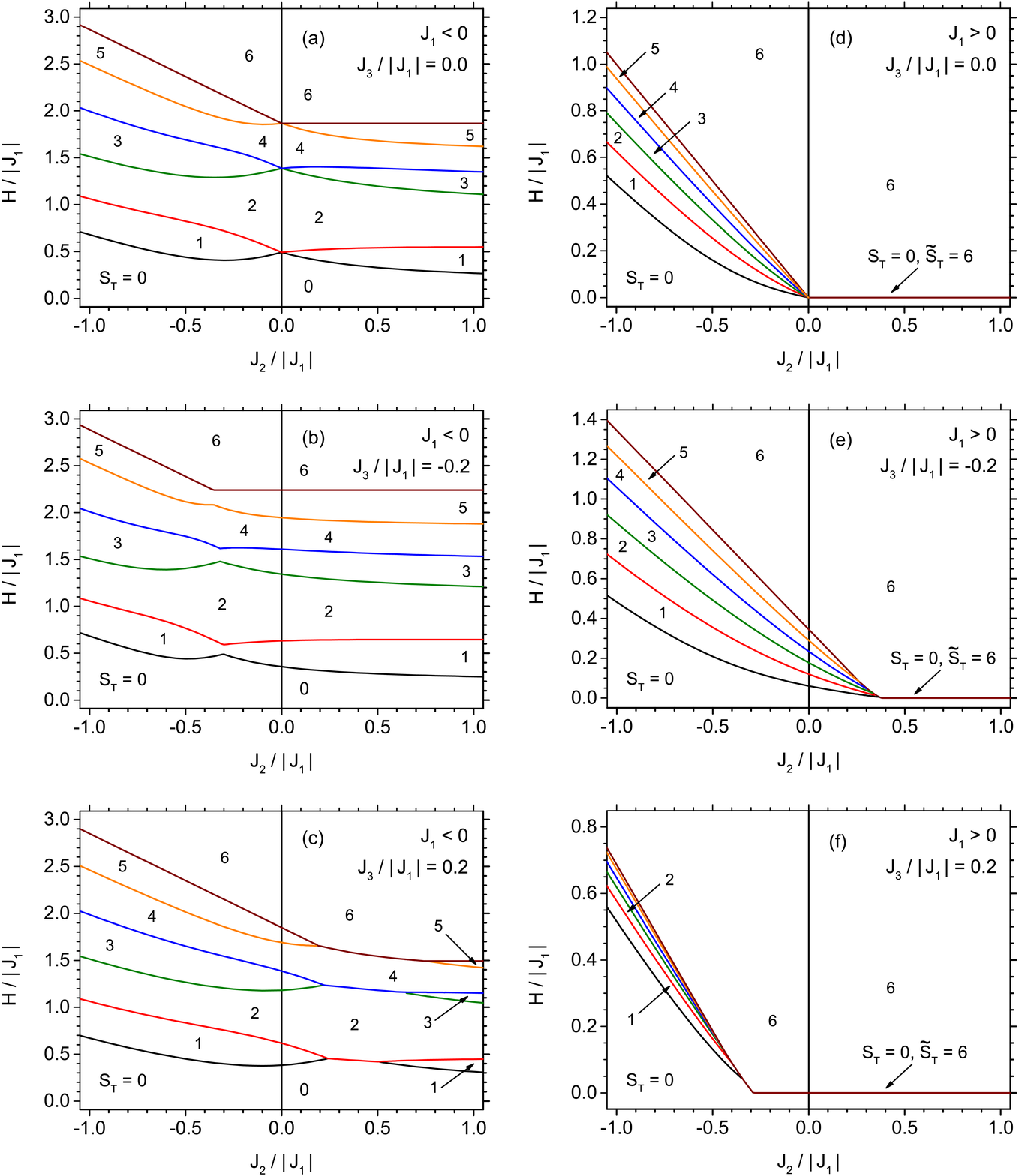}
  \end{center}
   \caption{\label{fig:2}ground-state magnetic phase diagram presenting the boundaries of the phases with various total spins (as indicated with the numbers), as a function of the normalized rung exchange integral $J_2/|J_1|$ and normalized magnetic field $H/|J_1|$. The plots (a)-(c are for $J_1<0$, the plots (d)-(f) are for $J_1>0$. Various values of normalized exchange integral $J_3$ are used for calculations (see plots).}
\end{figure*}

\begin{figure}[h!]
  \begin{center}
   \includegraphics[width=0.75 \columnwidth]{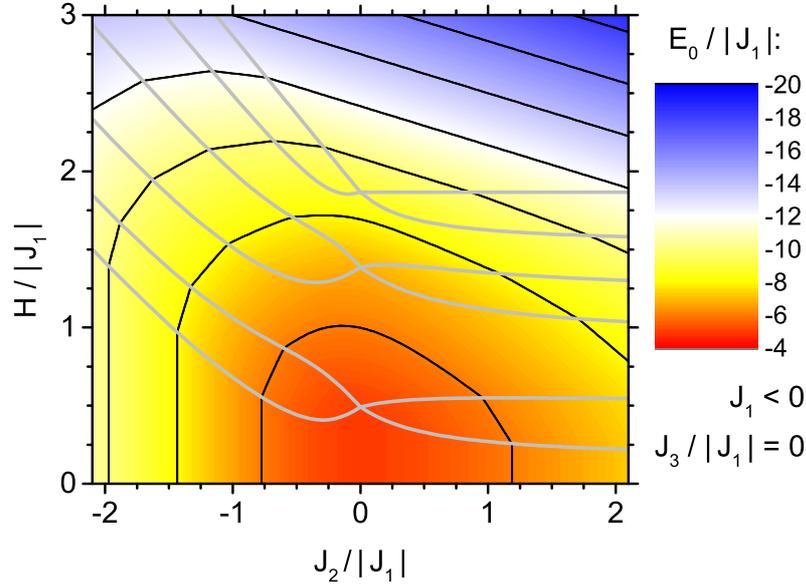}
  \end{center}
   \caption{\label{fig:en2}The contour plot of the normalized ground-state energy as a function of normalized external magnetic field $H/|J_1|$ and normalized rung exchange integral $J_2/|J_1|$. The bold lines correspond to constant normalized energy. The gray lines correspond to boundaries of  phases with various total spins [see Fig.~\ref{fig:2}(a)].  }
\end{figure}

\begin{figure}[h!]
  \begin{center}
   \includegraphics[width=0.75 \columnwidth]{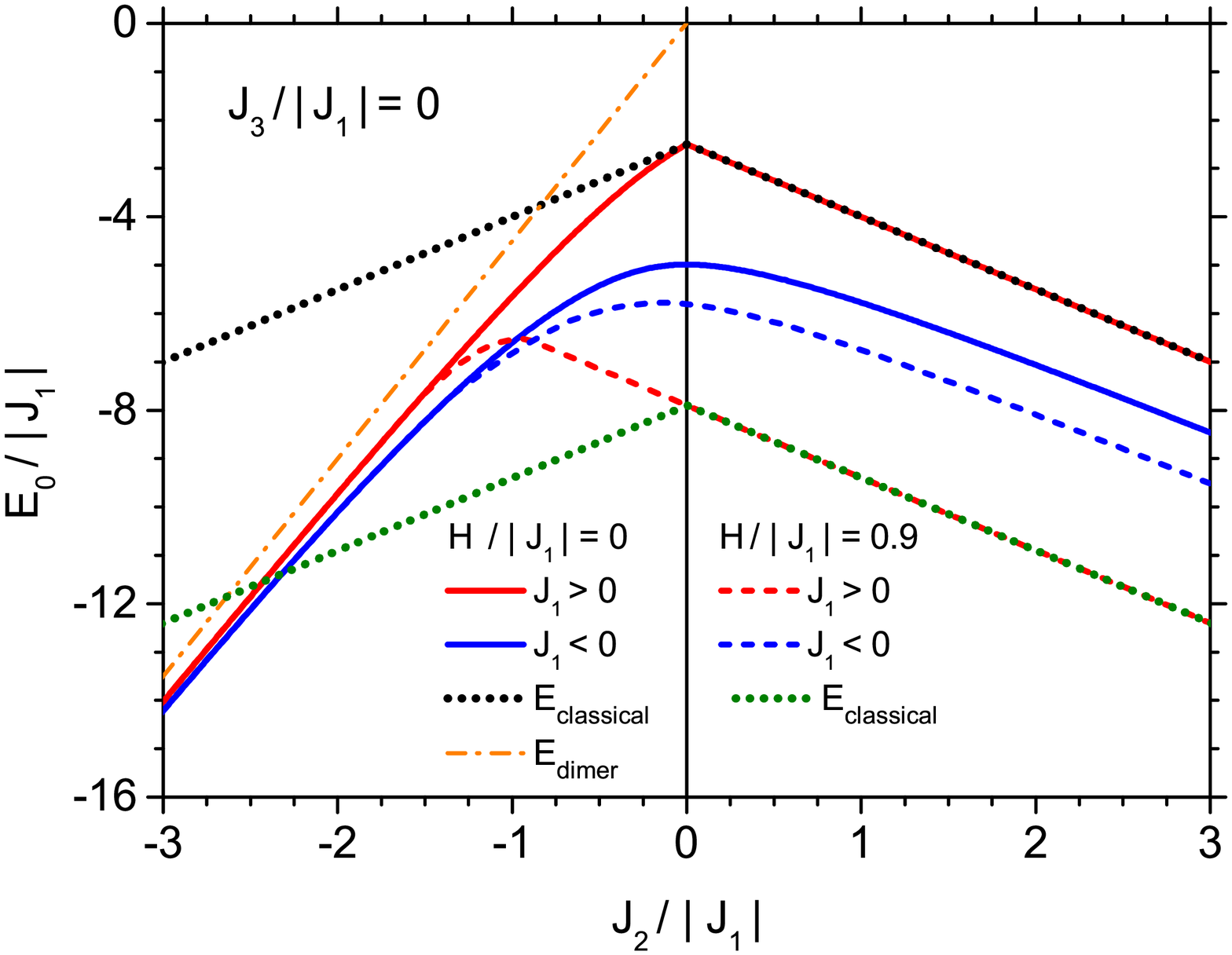}
  \end{center}
   \caption{\label{fig:en1}The normalized ground-state energy as a function of the normalized rung exchange integral $J_2/|J_1|$, for ferromagnetic and antiferromagnetic $J_1$ as well as for the magnetic fields $H/|J_1|=0.0$ and $H/|J_1|=0.9$. In addition, the ground-state energy of the classical state and the energy for the full dimerization of rungs are plotted.}
\end{figure}

All the calculations presented in this section rely on the exact numerical diagonalization of the system Hamiltonian (Eq.~(\ref{eq:1})) performed with the Mathematica software \cite{noauthor_mathematica_2010}. The discussion will be subdivided into subsections related to the key characteristics of the ground state.

\subsection{Ground-state phase diagram}

Let us commence the analysis from the investigation of the ground-state phase diagram for the system in the external magnetic field. The phases correspond here to various values of the $z$ component of the total spin, denoted by $S_T$. The phase diagram presenting the stability areas for phases with different values of $S_T$ as a function of $J_2/|J_1|$ and the magnetic field $H/|J_1|$ is shown in Fig.~\ref{fig:2}, for different values of $J_3/|J_1|$. The cases of antiferromagnetic $J_1<0$ and ferromagnetic $J_1>0$ are shown separately, as the absolute value $|J_1|$ is taken as the convenient energy to normalize other quantities. Let us analyse first the diagram for $J_1<0$ and $J_3/|J_1|=0.0$ [Fig.~\ref{fig:2}(a)]. At $J_2=0.0$ we deal with a pair of non-interacting finite spin chains, so the total spins take only even values. Introducing a finite, non-zero interaction $J_2$ restores the presence of the phases with all possible spins. What is interesting, for strongly antiferromagnetic $J_2$ the phase boundaries linearize and have identical slopes. This can be explained due to the fact that in the limit of dominant $J_2$ coupling the critical fields do not depend on $J_1$ (but only on $J_2$ itself). On the other hand, for strongly ferromagnetic $J_2$ the critical fields $H/|J_1|$ cease to depend on $J_2$. Let us mention that the diagram bears some resemblance to Fig.~2 in Ref.~\cite{cabra_magnetization_1997}, where the two-legged finite ladder was studied with the aim of characterizing an infinite system. 

If the antiferromagnetic crossing inter-leg interaction $J_3$ is switched on [as shown in Fig.~\ref{fig:2}(b)], the ladder legs are no longer non-interacting for $J_2=0.0$. As a consequence, for the full range of exchange integrals $J_2$ we pass through all the states with spins $0,\dots, 6$ when the magnetic field increases. However, close to some critical values of $J_2$ (slightly decreasing with the considered spin) the phases with odd spins are suppressed. The limiting behaviour of the diagram for strongly ferromagnetic and strongly antiferromagnetic coupling $J_2$ remains similar to the case of $J_3=0.0$. Contrary to the case of antiferromagnetic $J_3$, for ferromagnetic value $J_3/|J_1|=0.2$ [Fig.~\ref{fig:2}(c)], the range with only even values of spin expands from points $J_2=0.0$ to some finite intervals observable for positive $J_2$ (with their widths slightly rising when the involved spins increase).   

The diagram is completely different for ferromagnetic interaction $J_1>0$. In the absence of crossing inter-leg interaction $J_3=0.0$, as shown in Fig.~\ref{fig:2}(d), the spin is equal to 6 for $H>0$ and ferromagnetic $J_2$ (a saturated ferromagnetic state). For $J_2<0$ the states with all the possible spins are present, but the saturation is reached at considerably weaker field than in the case of antiferromagnetic $J_1$. If the antiferromagnetic crossing coupling $J_3<0$ is added [Fig.~\ref{fig:2}(e)], the critical magnetic fields separating the phases with various spins tend to increase. The field ranges corresponding to stability of a given spin are also wider. Moreover, the value of $J_2$, above which $H>0$ switches from spin 0 directly to spin 6, is increased (and corresponds to $J_2>0$). The narrow range where $H>0$ causes the switching from spin 0 to spin 6 with exclusion of some intermediate values is also noticeable. On the contrary, if $J_3>0$, as shown in Fig.~\ref{fig:2}(f), the critical magnetic fields are reduced and the field ranges for stability of a given spin are narrow. The critical value of $J_2$ above which switching directly between spin 0 and spin 6 takes place when $H$ increases is shifted towards antiferromagnetic (negative) values. Like in the case of $J_3>0$, also here a narrow interval in which some of the total spin values are missing when the field increases can be observed.

Let us comment on the quantum state corresponding to all the ranges considered in the described phase diagram (Fig.~\ref{fig:2}). In all the cases when $S_T>0$ in $H>0$, the quantum number of $z$ component of total spin $S_T$ is equal to the quantum number of the total spin $\tilde{S}_T$. Moreover, the state is non-degenerate, as the lowest energy is achieved by a single state (other projections $S_T<\tilde{S}_T$ are energetically unfavourable in the presence of the field). The situation is quite different for $S_T=0$ state. Two cases can be separated here. The state with $S_T=0$ seen in the case of $H=0$ and $J_1>0$ (changing to $S_T=6$ for arbitrarily weak field $H>0$ - see for example the range of $J_2>0$ at Fig.~\ref{fig:2}(d)) corresponds to $\tilde{S}_T=6$ and bears 13-fold degeneracy. This is due to the fact that for $H=0$ all the $2\tilde{S}_T+1$ possible values of total $z$ spin component ($-\tilde{S}_T,\tilde{S}_T+1,\dots,\tilde{S}_T$) lead to the same energy. Moreover, since we deal with canonical ensemble at ground-state, all such states are equally probable, so that all the possible values of total spin $z$ component average to $S_T=0$. On the contrary, the state with $S_T=0$ stable for the finite interval of magnetic fields up to the finite first critical magnetic field corresponds to the non-degenerate state with $S_T=0$ and $\tilde{S}_T=0$. 

\subsection{Ground-state energy}

The minimization of the energy decides on the form of the ground state taken by the system. However, it can be interesting to analyse the ground-state energy itself as a function of model parameters. The total normalized ground-state energy is plotted as a function of normalized rung interaction $J_2/|J_1$ and normalized magnetic field $H/|J_1|$ in Fig.~\ref{fig:en2} (for antiferromagnetic $J_1<0$ and in the absence of crossing coupling $J_3$). In the plot, the colour scale is used to indicate the ground-state energy; moreover, contour lines of constant energy are shown. In addition to these contours, gray lines correspond to the boundaries between phases with various $S_T$ (see Fig.~\ref{fig:2}(a)). It can be noticed that the presence of the magnetic field in principle always decreases the ground-state energy. One exception is the phase with $S_T=0$, for which the total energy is $H$-independent. The discontinuities are seen in the slope of the constant energy lines at the phase boundaries. The energy can be observed to decrease much faster with absolute value of crossing coupling $|J_2|$ when $J_2<0$. 

The cross-sections of Fig.~\ref{fig:en2} at constant magnetic fields $H/|J_1|=0.0$ and $H/|J_1|=0.9$ can be followed in Fig.~\ref{fig:en1}. The data presented in the plot correspond to both signs on intraleg coupling, both antiferro- and ferromagnetic. Moreover, two kinds of reference energies are plotted. The first one is the optimized 'classical' ground-state energy, expressed by
\begin{equation}
E_{classical}=-\frac{1}{4}\left(10|J_1|+6|J_2|+10|J_3|\right)-\frac{1}{2}NH.
\end{equation} 
This is the minimum energy which can be achieved via the classical states, for which $\left\langle \mathbf{S}_i\cdot \mathbf{S}_j\right\rangle =1/4$. Such a correlation value is assumed for every pair of interacting spins to optimize the hypothetical ground-state energy (note that such construction would also minimize the energy of possibly frustrated bonds).

Another reference energy is the energy of a set of uncoupled Heisenberg dimers in their ground-state (singlet). Such state energy for a dimer with $\mathcal{H}_{dimer}=-J\mathbf{S_{i}}\cdot \mathbf{S_{j}}$  for $J<0$ amounts to $E_{singlet}=-3J/4$, with correlation $\left\langle \mathbf{S}_i\cdot \mathbf{S}_j\right\rangle = -3/4$; the correlation is isotropic in spin space. If $J_2$ is the dominant coupling, we can decompose the system into 6 dimers (each rung of the ladder is dimerized), having the total energy $E_{dimer}=6E_{singlet}=-9|J_2|/2$.

In Fig.~\ref{fig:en1} it can be seen that for ferromagnetic $J_1>0$ and $J_2>0$ the ground-state energy is equal to the classical energy $E_{classical}$. This is the trace of the classical ferromagnetic state. On the other side, for $J_2<0$ and $|J_2|\gg |J_1|$ the ground-state energy tends to $E_{dimer}$, indicating the tendency to forming a rung-dimerized ground state. In all the cases the energy for $J_1<0$ is smaller than for $J_1>0$. Moreover, the ground-state energy is less sensitive to external magnetic field for antiferromagnetic $J_1$ than for ferromagnetic $J_1$ coupling. If $H>0$ and $J_2<0$ for ferromagnetic $J_1>0$, the ground-state energy is some range of $J_2$ is higher that the optimum classical energy $E_{classical}$.  

\subsection{Spin-spin correlations}

\begin{figure}[h!]
  \begin{center}
   \includegraphics[width=0.75 \columnwidth]{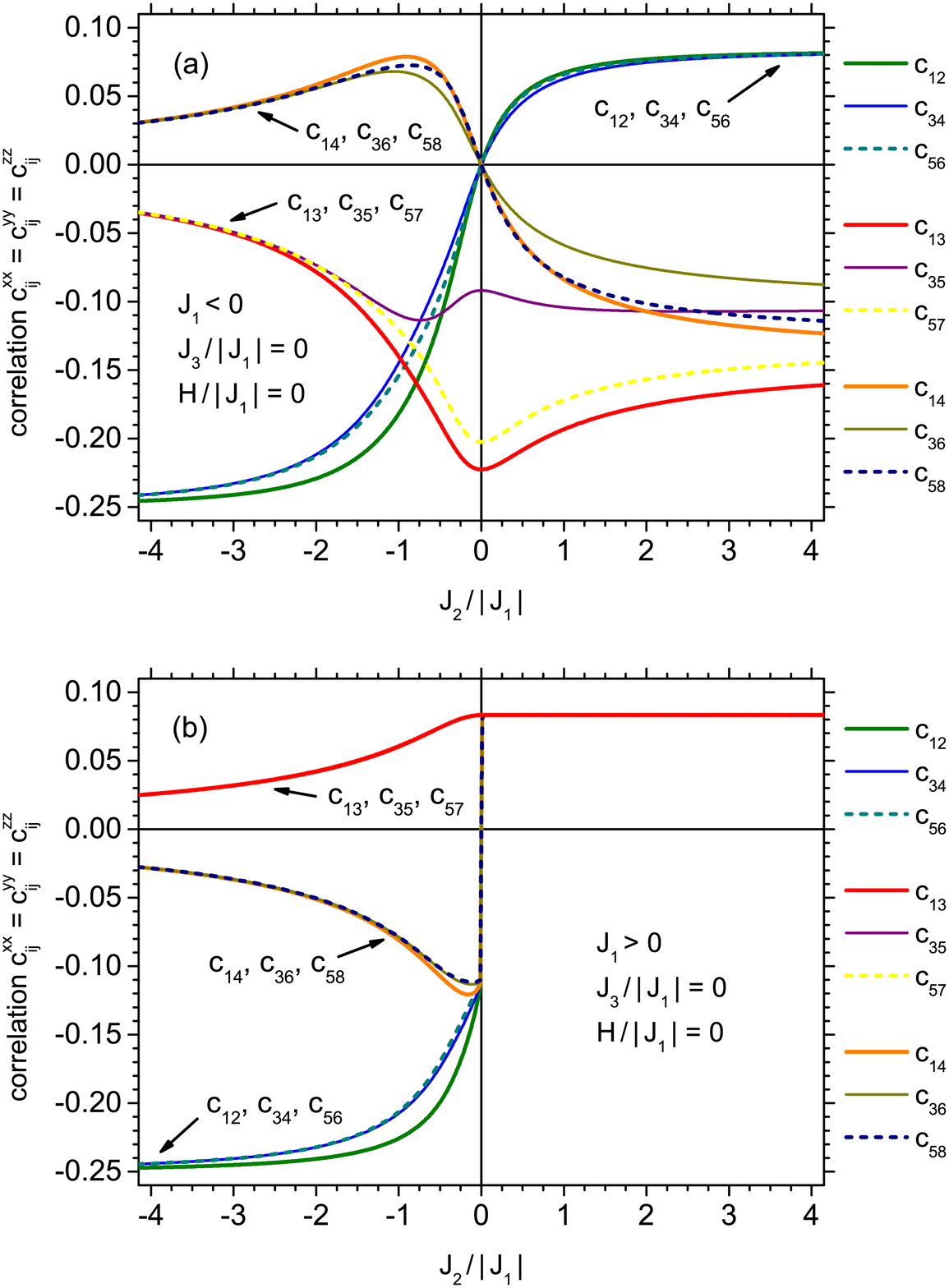}
  \end{center}
   \caption{\label{fig:corr1} The dependence of the isotropic spin-spin correlations on the normalized rung exchange integral $J_2/|J_1|$ in the absence of the magnetic field. For the explanation of spin pair labels see Fig.~\ref{fig:1}. (a) the case of $J_1<0$; (b) the case of $J_1>0$.  }
\end{figure}

\begin{figure}[h!]
  \begin{center}
   \includegraphics[width=0.75 \columnwidth]{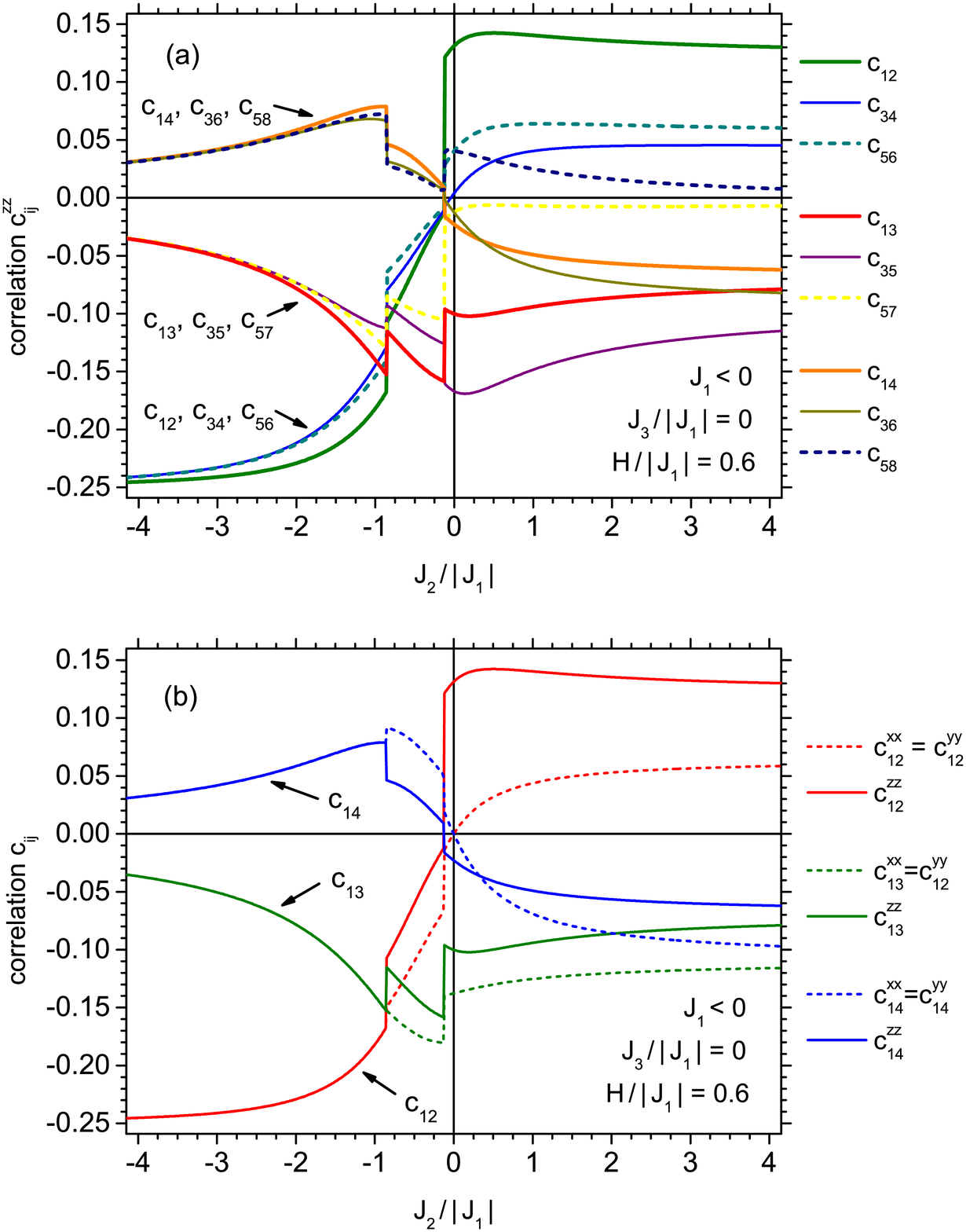}
  \end{center}
   \caption{\label{fig:corr2}The dependence of the spin-spin correlations on the normalized rung exchange integral $J_2/|J_1|$ in the presence of the magnetic field $H/|J_1|=0.6$, for the case of $J_1<0$. For the explanation of spin pair labels see Fig.~\ref{fig:1}. (a) All the correlations $c_{ij}^{zz}$; (b) the correlations $c_{ij}^{zz}$ and $c_{ij}^{xx}=c_{ij}^{yy}$ for the edge spin pairs.}
\end{figure}

The magnetic ground state of the quantum nanomagnet can be characterized in a more detailed way by calculating the spin-spin correlations. Since the considered nanostructure lacks translational symmetry, the correlations are site-dependent. Therefore, values of correlation functions for all the inequivalent spin pairs can be examined. An example of such selection of the correlation functions in the absence of the magnetic field, for the antiferromagnetic $J_1<0$, is shown in Fig.~\ref{fig:corr1}. The figure presents the inequivalent correlations between nearest neighbours in the same leg of the ladder, between interleg nearest-neighbours as well as crossing interleg correlations. The numerical indices label the pairs of sites at which the spins are located (for explanation see Fig.~\ref{eq:1}). It should be mentioned that for $H=0$ all the correlations are completely isotropic in spin space ($c_{ij}^{xx}=c_{ij}^{yy}=c_{ij}^{zz}$). In the limit of strongly antiferromagnetic $J_2$, all the correlations between nearest-neighbouring spins in different legs (rung correlations) tend to $-1/4$, as it is expected for quantum dimers (see also the discussion of the ground-state energy). Other correlations (between nearest neighbours in the same leg and crossing correlations) tend to vanish under the same conditions (the intraleg correlations are negative, while the interleg crossing correlations are positive). When $J_2$ is reduced in magnitude, the interleg correlations become weaker (and achieve $0$ at $J_2=0$), whereas the nearest-neighbour interleg correlations are peaked. In all the cases the correlations for inequivalent spin pairs differ (which is a finite size effect in the nanoscopic system), however, the effect becomes particularly strong  for weak $J_2$. A trace of dimerization behaviour within each leg can be noticed for weak $J_2$, as every second correlation is significantly smaller. On the ferromagnetic side of $J_2$, the rung correlations tend to $1/12$ (and the sum of correlations $c_{ij}^{xx}=c_{ij}^{yy}=c_{ij}^{zz}=\left\langle\mathbf{S}_{i}\cdot \mathbf{S}_{j}\right\rangle$ amounts to $1/4$ as for the classical ferromagnetic state). Other correlations are negative and tend slowly to some limiting values.

The behaviour of correlations is completely different for the case of ferromagnetic $J_1>0$, as shown in Fig.~\ref{fig:corr1}(b). There, for $J_2<0$, the rung correlations again tend to $-1/4$ when $J_2$ becomes dominant, and other correlations tend to vanish. For ferromagnetic $J_2$, all the correlations take the common value of $1/12$, which is further independent on $J_2$. At crossing the value of $J_2=0$ the interleg correlations behave discontinuously, while the intraleg nearest-neighbours correlation varies in continuous manner. For vanishing $J_2$ from the antiferromagnetic side, all the interleg correlations take the common value. In the case of $J_1>0$ the differences between the correlations for inequivalent pairs of spins are much less pronounced than in the case of $J_1<0$.

Let us remind that in the absence of the magnetic field we deal with the states characterized by $S_T=0$ and either $\tilde{S}_T=6$ (which corresponds to $J_2>0$ in Fig.~\ref{fig:corr1}(b)) or $\tilde{S}_T=0$  (for the rest of Fig.~\ref{fig:corr1}).

The introduction of the external magnetic field exerts an important effect on the behaviour of the correlations. The most important influence is inducing the spin-space anisotropy. For the case of $J_1<0$ the effect of the magnetic field can be traced in Fig.~\ref{fig:corr2} (this corresponds to the cross-section of the phase diagram presented in Fig.~\ref{fig:2}(a)). The part Fig.~\ref{fig:corr2}(a) shows the inequivalent correlations of the type $c_{ij}^{zz}$ as a function of normalized $J_2/|J_1|$. For the most negative values of $J_2$ we deal with the state $S_T=0$ (see Fig.~\ref{fig:2}(a)) and the behaviour of the correlations resembles the case from Fig.~\ref{fig:corr1}(a), close to dimerization with respect to ladder rungs. For weakly negative $J_2$ we deal with a cross-over to the state $S_T=1$, and further increase in $J_2$ causes the state $S_T=2$ to be achieved. In the latter state, in can be seen that the intraleg nearest-neighbour correlations in the middle of the ladder are strongly suppressed, while similar correlations closer to the edges take more pronounced (negative) values. The situation is somehow similar for the rung correlations and crossing correlations. In general, states with higher spins characterize themselves with more significant differences between correlations for inequivalent spin pairs of the same type.

The spin-space anisotropy in correlations can be followed in Fig.~\ref{fig:corr2}(b), where the correlations of all three types are shown for spin pairs close to the edge. In the state $S_T=0$ the correlations are isotropic, while for $S_T>0$ a difference between $c_{ij}^{zz}$ and $c_{ij}^{xx}=c_{ij}^{yy}$ emerges. It is interesting that for $S_T=1$ the absolute values of $zz$ correlations are smaller than the $xx$ and $yy$ correlations. On the contrary, for $S_T=2$, the (positive) $zz$ rung correlations dominate over $xx$ and $yy$ components, while the remaining correlations are negative and $zz$ components are weaker than the other in that case.

\begin{figure}[h!]
  \begin{center}
   \includegraphics[width=0.75 \columnwidth]{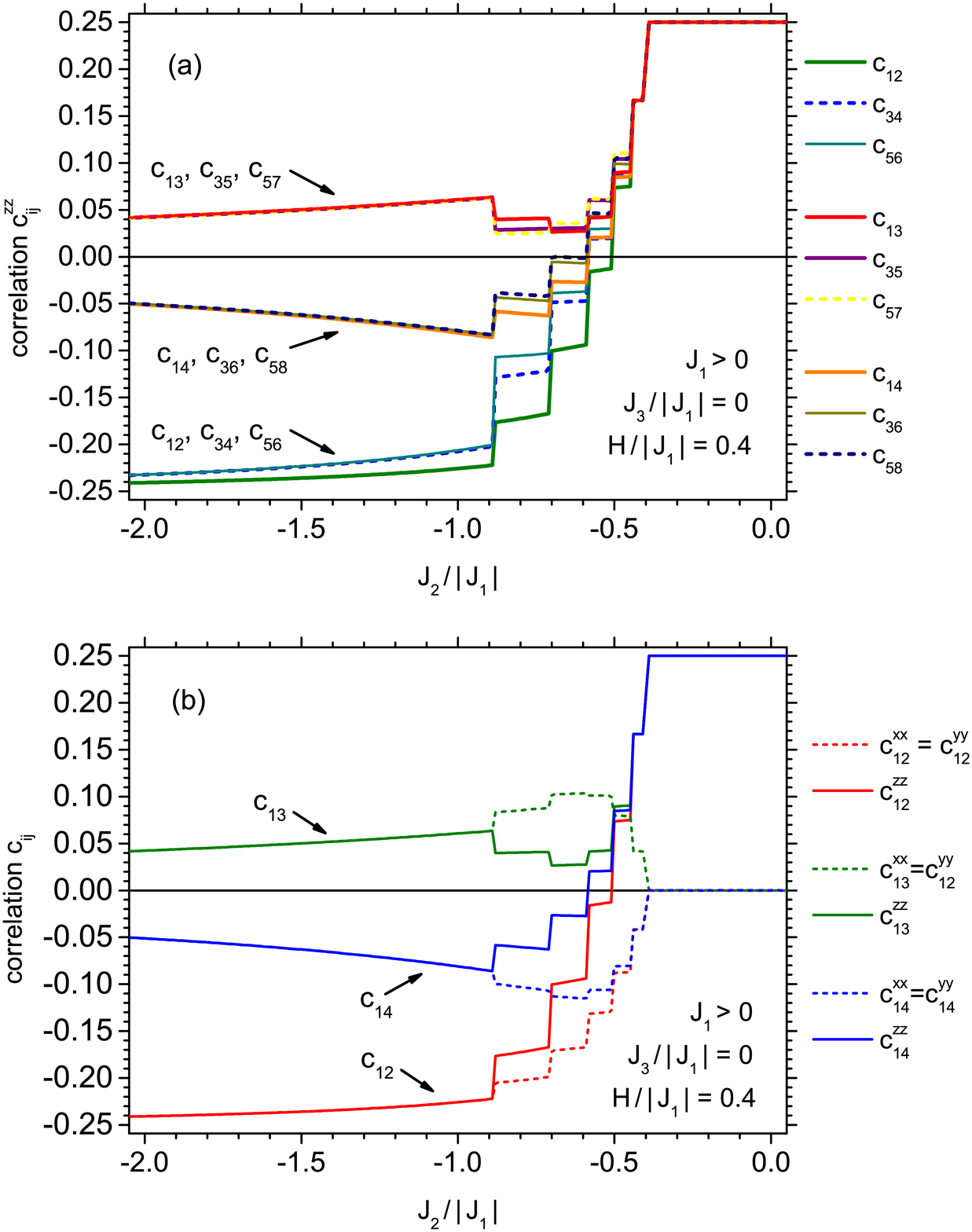}
  \end{center}
   \caption{\label{fig:corr3}The dependence of the spin-spin correlations on the normalized rung exchange integral $J_2/|J_1|$ in the presence of the magnetic field $H/|J_1|=0.4$, for the case of $J_1>0$. For the explanation of spin pair labels see Fig.~\ref{fig:1}. (a) all the correlations $c_{ij}^{zz}$; (b) the correlations $c_{ij}^{zz}$ and $c_{ij}^{xx}=c_{ij}^{yy}$ for the edge spin pairs. }
\end{figure}

\begin{figure}[h!]
  \begin{center}
   \includegraphics[width=0.75 \columnwidth]{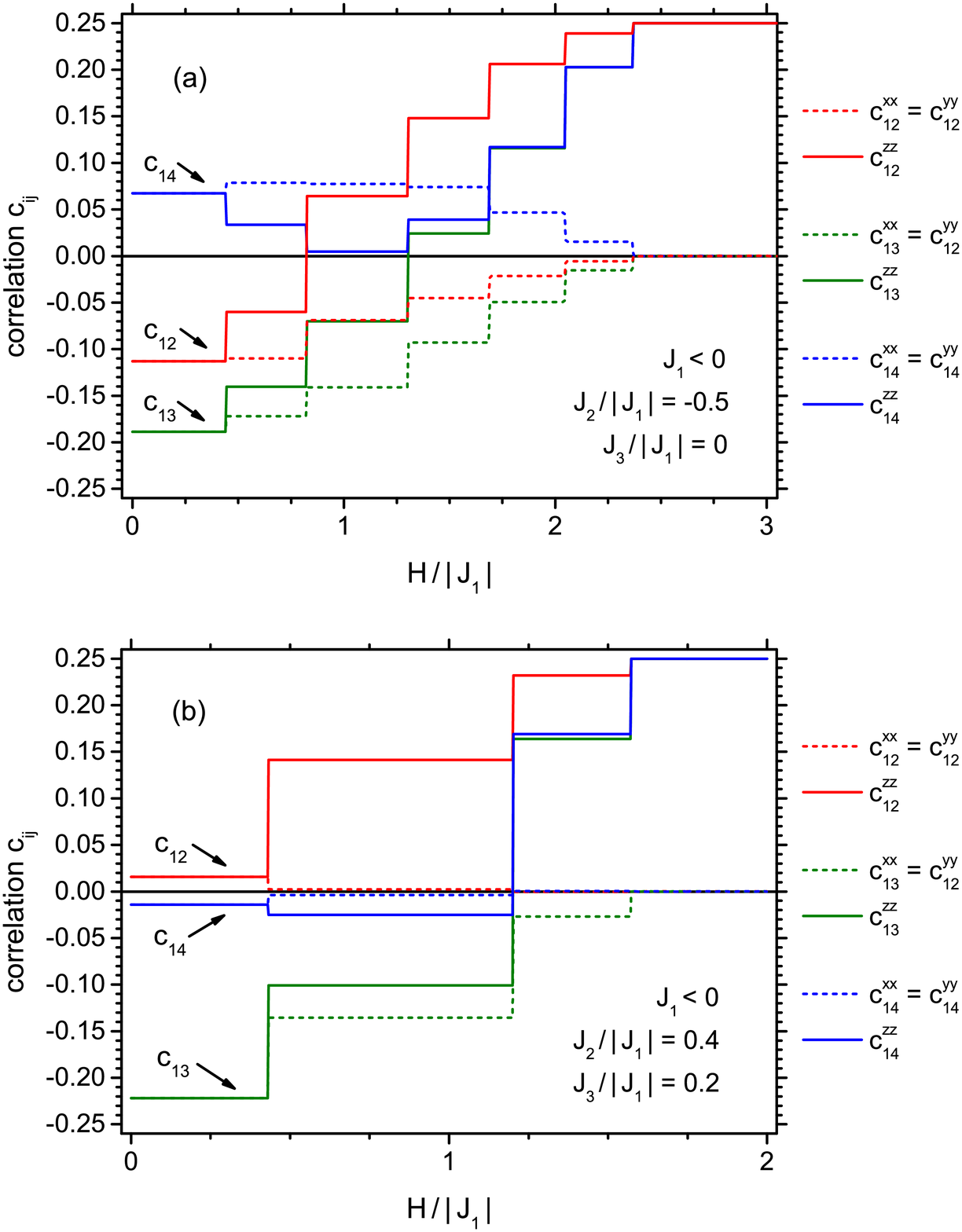}
  \end{center}
   \caption{\label{fig:corr4}The dependence of the spin-spin correlations $c_{ij}^{zz}$ and $c_{ij}^{xx}=c_{ij}^{yy}$ on the normalized magnetic field $H/|J_1|$ for the case of $J_1<0$. For the explanation of spin pair labels see Fig.~\ref{fig:1}. (a) the case of $J_2/|J_1|=-0.5$ and $J_3/|J_1|=0.0$; (b) the case of $J_2/|J_1|=0.4$ and $J_3/|J_1|=0.2$ }
\end{figure}

The effect of the magnetic field on the correlations for the case of $J_1>0$ can be traced in Fig.~\ref{fig:corr3}. As it is seen in the corresponding phase diagram (Fig.~\ref{fig:2}(d)), increase in $J_2$ causes crossing of the borders between all the states starting from $S_T=0$ up to $S_T=6$. Once more, for strong rung interactions, the correlations resemble the case presented in Fig.~\ref{fig:corr1}(b). The rung correlations (negative) dominate and tend gradually to switch to positive (ferromagnetic) character when $J_2$ increases. The remaining correlations show slightly different behaviour. For $S_T\geq 4$ all the correlations become positive. What is interesting, for $S_T\geq 5$ all the correlations take the common values (and for $S_T=6$ this value is $1/4$, corresponding to the classical saturated ferromagnetic state). For $S_T=1,\dots,4$ the differences between correlations for various inequivalent locations of spin pairs of the same type are more pronounced. Fig.~\ref{fig:corr3}(b) allows the analysis of the spins pace anisotropies developing under the influence of the magnetic field. For $S_T\geq 5$ all the $zz$ components take the common positive values, whereas $xx$ and $yy$ components vanish. For the intermediate cases of $S_T=1,2,3$ the $zz$ components of all correlations are weaker in magnitude than $zz$ components.

The dependence of correlations for spin pairs close to the edge on the normalized magnetic field is presented for two selections of interaction parameters in Fig.~\ref{fig:corr4}, in both cases for $J_1<0$. In general, it can be seen that the correlations change in step-wise manner, so each state with given $S_T$ corresponds to one plateau. In Fig.~\ref{fig:corr4}(a) a cross-section of the phase diagram Fig.~\ref{fig:2}(a) reveals that all the intermediate states between $S_T=0$ and $S_t=6$ are achieved when $H$ rises. The intraleg nearest-neighbour correlations and rung correlations increase up to saturation, while the crossing correlations first tend to be reduced in their magnitude and then increase. In the case of nearest-neighbour intraleg correlations and rung correlations the $zz$ component can be positive when $xx$ and $yy$ component take negative values, proving the pronounced anisotropy in spin space.

In Fig.~\ref{fig:corr4}(b) the selection of the interaction parameters corresponds to such cross-section of the phase diagram (Fig.~\ref{fig:2}(c)) that only the states with even $S_T$ are crossed when $H$ increases. This is reflected in the number of correlations plateaux. The qualitative behaviour of the correlations resembles the previous case. However, this time the only positive correlations at weak field are the rung correlations. Moreover, at weak field only the nearest neighbour intraleg correlations take the significant values.

\section{Final remarks}
\label{Final remarks}

The calculations presented above constitute an extensive analysis of the ground-state behaviour of the selected magnetic nanocluster, of the shape of two-legged finite ladder consisting of 12 spins $S=1/2$. The interactions in the system corresponded to isotropic quantum Heisenberg model. The obtained results originate from exact numerical diagonalization of such model and provide a view free from approximation artefacts. An influence of the external magnetic field and of the rung coupling on the total spin was illustrated in a phase diagram, calculated for both positive and negative intraleg coupling and for various crossing interactions. The ground-state energy was analysed as a function of the rung coupling and magnetic field. Moreover, the behaviour of the spin-spin correlations for all inequivalent spin pairs of the same type was characterized as a function of the model parameters. In particular, the emergence of spin-space anisotropies under the action of magnetic field was analysed. The comparison of the cases of ferromagnetic and antiferromagnetic intraleg nearest-neighbour couplings was performed, yielding crucial differences between both classes of nanomagnets.

The obtained results show an interesting behaviour and a variety of quantum phases exhibited by the studied nanomagnet, demonstrating the large degree of tunability. The possible extensions of the present work involve mainly the consideration of the thermal properties. Moreover, the cases of higher spins can be studied, allowing inclusion of additional interactions and other relevant Hamiltonian terms.

\section*{Acknowledgments}

\noindent This work has been supported by Polish Ministry of Science and Higher Education on a special purpose grant to fund the research and development activities and tasks associated with them, serving the development of young
scientists and doctoral students.


\end{document}